\documentclass[aps,twocolumn,nofootinbib]{revtex4-2}
\usepackage[utf8]{inputenc}
\usepackage{amsmath,bm}
\usepackage{amsfonts}
\usepackage{amssymb}
\usepackage{xcolor}
\usepackage{braket}
\usepackage{graphicx}
\usepackage{subfigure}
\usepackage{todonotes}
\usepackage{subfigure}
\usepackage[colorlinks=true, allcolors=blue]{hyperref}
\usepackage{dsfont}

\usepackage{etoolbox} 
\usepackage{lipsum} 
\usepackage[capitalize]{cleveref}

\begin{document}

\title{Detecting nonclassicality in randomly-displaced copies of a squeezed state}

\author{Mehmet Emre Tasgin}
\affiliation{Institute of Nuclear Sciences, Hacettepe University, 06800 Ankara, Turkey}

\date{\today}

\begin{abstract}
	We address a fundamental question: Can one determine whether a received signal is squeezed when each copy arrives with a different displacement/amplitude?
	We introduce an interaction Hamiltonian that converts quadrature squeezing into number squeezing. Using this conversion, we test whether the copies satisfy $g^{(2)}(0)<1$. The Hamiltonian itself does not create nonclassicality; it only transfers it from quadrature squeezing to number squeezing. This allows us to identify squeezing even when individual copies have random displacements.
\end{abstract}

\maketitle

\section{Introduction}

Squeezed light is an indispensable resource for quantum technologies~\cite{dowling2003quantum,o2009photonic}. It can be converted into different forms of entanglement using beam splitters (BS)~\cite{ge2015conservation} or BS-like interactions~\cite{kuzmich1997spin,tacsgin2011spin}. For example, the output modes of a BS become entangled only if the input mode is nonclassical (such as squeezed)~\cite{kim2002entanglement,asboth2005computable}. In this case, a conservation-like relation holds between squeezing and entanglement~\cite{ge2015conservation,liu2024classical,tasgin2020quantifications,arkhipov2016interplay}. Quadrature squeezing can also be converted into many-particle entanglement through a BS-like interaction with atoms~\cite{kuzmich1997spin,tacsgin2011spin}.

Experimental determination of squeezing requires millions of identically prepared copies of the state~\cite{wu1986generation,park2024single}. In each copy, the quadrature displacement value $\alpha$ must remain the same in order to obtain the statistical noise distribution around that value. Environmental effects, however, may change the displacement value of successive copies. For example, the atmosphere is a dynamical channel that can produce different displacements for consecutive copies due to varying absorption and scattering~\cite{karsa2024quantum,oruganti2025continuous,fesquet2023perspectives,barzanjeh2011entangling}. Another example appears in quantum imaging of biological samples~\cite{li2024harnessing,taylor2013biological}, where light undergoes multiple scattering events and the displacement becomes randomized.

In some continuous-variable quantum key distribution protocols employing squeezed states, it may be relevant for Bob to assess whether squeezing is preserved, for example for channel characterization or performance benchmarking~\cite{oruganti2025continuous,fesquet2023perspectives,cerf2001quantum}. Moreover, in a recent quantum communication method~\cite{tasgin2025encoding}, detecting whether the received light is squeezed is crucial because the classical information is encoded in the squeezing degree of freedom.

Therefore, witnessing the nonclassicality of an originally squeezed state in a dynamical or uncontrolled-scattering environment is a major challenge.

In this paper, we develop a method for witnessing the nonclassicality of a randomly displaced squeezed (RDS) light source. Each received copy of the originally squeezed state has a different and unknown displacement. Because of this, the presence of squeezing cannot be detected using standard quadrature-measurement techniques~\cite{wu1986generation,park2024single,gardiner2004quantum}. To overcome this problem, we design an interaction Hamiltonian that converts quadrature squeezing into number squeezing. We then witness the nonclassicality through the $g^{(2)}(\tau)$ histogram of the converted mode. We show that this interaction Hamiltonian does not generate any nonclassicality unless the received copy is originally squeezed.

\begin{figure*}
	\centering
	\includegraphics[width=1.00 \textwidth]{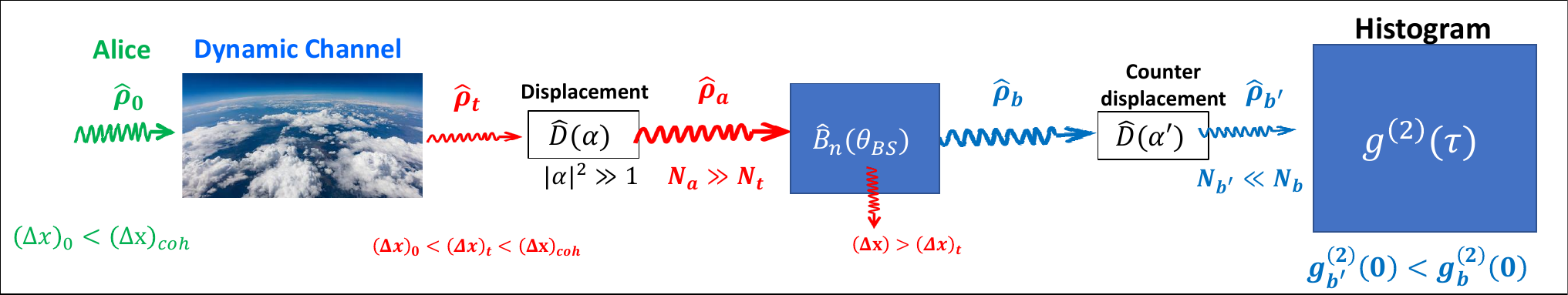}
	\caption{{\it Protocol for detecting nonclassicality in randomly displaced copies of a squeezed state.}
		(i) Alice sends identical copies of a squeezed state $\hat{\rho}_0$, but each copy reaches the receiver as $\hat{\rho}_t$ with an unknown displacement due to attenuation in the dynamical channel.
		(ii) The received copy is displaced by $\hat{D}(\alpha)$ with a large $\alpha$, i.e., $|\alpha|^2 \gg 1$, to obtain $\hat{\rho}_a$. The operation $\hat{D}(\alpha)$ does not change the nonclassicality of $\hat{\rho}_t$~\cite{kim2002entanglement,simon1994quantum,adesso2010quantum,asboth2005computable}, but the resulting states $\hat{\rho}_a$ have much narrower pulse-to-pulse variations in their mean photon number.
		(iii) $\hat{\rho}_a$ enters the $\hat{B}_n(\theta)$ operation, which converts the quadrature squeezing in $\hat{\rho}_a$ into number squeezing in the output state $\hat{\rho}_b$ (see Fig.~\ref{fig2}a). As a result, the quadrature squeezing in the $\hat{a}$-mode output decreases (see Fig.~\ref{fig2}b). A strong coherent state is used as the input for the $\hat{b}_n$ port.
		(iv) Although the number squeezing in $\hat{\rho}_b$ is well below unity, the large photon number keeps $g^{(2)}(0)$ close to 1. To make $g^{(2)}(0)<1$ more visible, we apply a counter-displacement $\hat{D}(\alpha')$. (See Fig.~\ref{fig3}b for how much $g^{(2)}(0)$ can be reduced even when its initial value is 0.995.)
		(v) The resulting state $\hat{\rho}_{b'}$ is then used to build the $g^{(2)}(\tau)$ histogram over many received pulses.
		(vi) Post-selection can also be applied to further reduce pulse-to-pulse photon-number variations, allowing clearer visualization of $g^{(2)}(0)<1$.
	}
	\label{fig1}
\end{figure*}

Let us describe the steps of our detection protocol one by one in Fig.~\ref{fig1}.
We receive squeezed states with random and attenuated amplitudes (displacements), i.e.,
$|\alpha_1,r_1\rangle$, $|\alpha_2,r_2\rangle$, $|\alpha_3,r_3\rangle$, $\ldots$ — represented as $\hat{\rho}_t$ in Fig.~\ref{fig1}.
The squeezing rates $r_i$ also degrade and can be different for each copy.
We apply the same displacement operation
$\hat{D}(\alpha)=\exp( \alpha \hat{a}^\dagger - \alpha^* \hat{a} )$
with $|\alpha|^2 \gg |\alpha_i|^2$ to each copy.
The operation $\hat{D}(\alpha)$ does not change the nonclassicality of the state~\cite{RSimon94,kim2002entanglement,simon1994quantum,asboth2005computable,adesso2010quantum,GAdesso04,serafini2003symplectic}.
We denote the state after this displacement as $\hat{\rho}_a$.

This displacement is performed for two reasons.
First, we want to reduce the pulse-to-pulse variation of the mean photon number $\langle\hat{n}_i\rangle$ among different copies.
This makes it easier to detect $g^{(2)}(0)$ in the histogram.
Second, our analytical results showing how the interaction Hamiltonian converts quadrature squeezing into number squeezing hold in the large $\langle\hat{n}\rangle$ regime.

Next, we introduce the pronounced beam-splitter (BS)-like interaction
\begin{equation}
	\hat{B}_{n}(\theta)= \exp\left[ \theta (e^{i\phi} \hat{b}_n^\dagger \hat{a} - e^{-i\phi}  \hat{a}^\dagger \hat{b}_n) \right]
\end{equation}
which is analogous to the usual BS operator $\hat{B}(\theta)$~\cite{Scully_Zubairy_1997}.
(The operator $\hat{B}_n(\theta)$ is studied in detail in Sec.~\ref{sec:BSn}.)
Here, instead of the usual annihilation operator $\hat{b}$, we use the analogous operator 
\begin{equation}
	\hat{b}_n=(\hat{n} + i \gamma_0 \hat{\Phi})/\sqrt{2\gamma_0} \: ,  \quad \gamma_0=2\langle\hat{n}\rangle,
\end{equation}
whose properties were studied in Refs.~\cite{vaccaro1990physical,tasgin2019anatomy,pegg1989phase,barnett1989hermitian,barnett2007quantum}.
We also analyze $\hat{b}_n$ in Sec.~\ref{sec:bn_operator}.
Since the canonical variables associated with $\hat{b}_n$ are number and phase,
$\hat{B}_n(\theta)$ transforms quadrature squeezing
($\hat{a}=(\hat{x}+i\hat{p})/\sqrt{2}$)
into number squeezing
($\hat{b}_n= (\hat{n} + i\gamma_0 \hat{\Phi})/\sqrt{2\gamma_0}$).

The state after this conversion is denoted by $\hat{\rho}_b$ in Fig.~\ref{fig1}. The second output port of the $\hat{B}_n(\theta)$ operation (red vertical arrow) carries reduced quadrature squeezing. We show that the increase in nonclassicality of the $\hat{b}_n$ mode in Fig.~\ref{fig2}a is accompanied by a corresponding decrease of nonclassicality in the red output mode, i.e., $(\Delta \hat{x}) > (\Delta \hat{x})_t$, as demonstrated in Fig.~\ref{fig2}b through our numerical simulations (Sec.~\ref{sec:numeric}).

We go further and demonstrate the transfer of quadrature squeezing into number squeezing analytically in Sec.~\ref{sec:SqzTransfer}. To this end, we introduce in Sec.~\ref{sec:AnalSqz} the squeezing operator $\hat{S}_n(\xi)$, which is the analogue of the usual squeezing operator~\cite{ScullyZubairyBook} for the $\hat{b}_n$ mode. The analytical predictions are in full agreement with the numerical simulations shown in Fig.~\ref{fig2}. For completeness, we also define in Sec.~\ref{sec:AnalogDisp} the corresponding displacement operator, which generates displacements in the number–phase plane, analogous to displacements in the $x$–$p$ phase space.

The most crucial point is shown in Sec.~\ref{sec:noNc}.
There we prove that $\hat{B}_n(\theta)$ cannot generate nonclassicality unless it is already present in the $\hat{a}$ input mode.
The proof is analogous to Ref.~\cite{kim2002entanglement} for an ordinary BS.
This means that the $\hat{b}_n$ mode can show nonclassicality with $g^{(2)}(0)<1$ only if the incoming randomly displaced copy is originally squeezed.

Before sending $\hat{\rho}_b$ into the $g^{(2)}(\tau)$ histogram, we address another issue (Sec.~\ref{sec:counterDisplacement}).
The number squeezing $(\Delta \hat{n}_i)^2 / \langle \hat{n}_i\rangle$ in $\hat{\rho}_b$ becomes much smaller than unity after the $\hat{B}_n(\theta)$ transformation.
However, we cannot measure number squeezing directly.
We only measure $g^{(2)}(0)$, which remains close to unity because of the large photon number, e.g., $0.98$ in Fig.~\ref{fig2}a.
Therefore, before histogramming the state, we apply a counter-displacement operator $\hat{D}(\alpha')$ that reduces the mean photon number of $\hat{\rho}_b$.
The operation $\hat{D}(\alpha')$ does not create nonclassicality unless it is already present~\cite{RSimon94,kim2002entanglement,adesso2010quantum,asboth2005computable}.

Figure~\ref{fig3} shows that this counter-displacement—which reduces the mean photon number (Fig.~\ref{fig3}a)—enhances the visibility of the $g^{(2)}(0)$ dip (Fig.~\ref{fig3}b), at the expense of reduced number squeezing (Fig.~\ref{fig3}c). In Fig.~\ref{fig3}, we deliberately choose an example with an initial value $g^{(2)}(0)\simeq 0.995$ to demonstrate that even values extremely close to unity can be made clearly visible in a histogram. We note that the minimum value of $g^{(2)}(0)$ obtained in Fig.~\ref{fig2}a is already significantly smaller—hence more visible—even without counter-displacement. Applying the same counter-displacement to this minimum value ($g^{(2)}(0)=0.98$ in Fig.~\ref{fig2}a) would therefore yield even lower observable values of $g^{(2)}(0)$.

In Sec.~\ref{sec:histogram}, we describe in detail how $g^{(2)}(\tau)$ is measured experimentally. We discuss the robustness of $g^{(2)}(\tau)$ against pulse-to-pulse fluctuations in the mean photon number, which are common in practice. We show that these fluctuations are naturally suppressed in our scheme and that the visibility of sub-Poissonian statistics can be further enhanced by post-selection performed after the counter-displacement, enabling clearer observation of $g^{(2)}(0)<1$.

We note that, in the counter-displacement operation $\hat{D}(\alpha')$, we do not reduce the mean photon number $N_{b'}^{(i)}$ in Fig.~\ref{fig1} too much. Specifically, we keep $N_{b'}^{(i)} \gg \langle \hat{n}_i\rangle$. As a result, the relative number fluctuations among the $N_{b'}^{(i)}$ values are orders of magnitude smaller than those of the incident $\langle \hat{n}_i\rangle$. If $N_{b'}^{(i)}$ were reduced too strongly in the counter displacement $\hat{D}(\alpha')$, the relative number fluctuations would become comparable to those of the output states $\hat{\rho}_b^{(i)}$ (or $\hat{\rho}_t^{(i)}$).

To further suppress relative fluctuations, we employ a simple technique—described as method 3 in Sec.~\ref{sec:histogram}—in which all $N_{b'}^{(i)}$ values are scaled by a common factor. This reduces the effect of $\langle \hat{n}_i\rangle$ fluctuations by several orders of magnitude, bringing them down to levels typical of conventional single-photon detection experiments~\cite{loudon2000quantum,fox2006quantum,migdall2013single,lounis2005single}. This scaling is implemented by passing the $N_{b'}^{(i)}$ fields through an attenuating medium, where the amplitudes decay exponentially by the same factor for all $N_{b'}^{(i)}$, while the quantities $g_i^{(2)}(0)$ remain unchanged. See Refs.~\cite{di2012quantum,lopaeva2013experimental,sekatski2012detector} and the derivation in  Appendix~\ref{sec:appendix_attenuation}.



We present the details on our numerical simulations in Sec.~\ref{sec:numeric}.
Finally, in Sec.~\ref{sec:conclusions}, we summarize our results and discuss their implementation. The physical systems that can realize the $\hat{B}_n(\theta)$ Hamiltonian are presented in Appendix~\ref{sec:appendix_Hamiltonian}.

\section{Number-phase Minimum Uncertainty States} \label{sec:bn_operator}

In this section, we introduce key properties of minimum-uncertainty (intelligent) number–phase states and their associated operators~\cite{vaccaro1990physical,aragone1976intelligent,tasgin2019anatomy,pegg1989phase,barnett1989hermitian,barnett2007quantum}. These concepts will be used extensively throughout the paper.

The algebra of the standard annihilation operator $\hat{a} = (\hat{x} + i\hat{p})/\sqrt{2}$ is well established in quantum mechanics~\cite{sakurai1967advanced} and quantum optics~\cite{Scully_Zubairy_1997}. Its eigenstates are coherent states, which lie at the boundary between classical and nonclassical regimes~\cite{drummond1980generalised,kiesel2010nonclassicality}. When the operator is generalized to $\hat{a}(s) = (\hat{x} + i s \hat{p})/\sqrt{2}$, its eigenstates correspond to quadrature-squeezed states: $x$-squeezed for $s<1$ and $p$-squeezed for $s>1$. These states satisfy the minimum-uncertainty relation
\begin{equation}
	\langle(\Delta \hat{x})^2\rangle \,\langle(\Delta \hat{p})^2\rangle = \frac{1}{4}.
\end{equation}
Analogously, one can define the operator~\cite{vaccaro1990physical,tasgin2019anatomy,pegg1989phase,barnett1989hermitian,barnett2007quantum}
\begin{equation}
	\hat{b}_n=\frac{\hat{n}+ i \gamma_0 \hat{\Phi}}{\sqrt{2\gamma_0}},
	\label{bn_operator}
\end{equation}
constructed from the canonically conjugate pair $\hat{n}$ and $\hat{\Phi}$, with $\gamma_0 = 2\langle \hat{n} \rangle$. In the limit $\langle\hat{n}\rangle\gg 1$, these operators satisfy a commutation relation analogous to that of $\hat{x}$ and $\hat{p}$, namely $[\hat{n},\hat{\Phi}] = i$~\cite{vaccaro1990physical,pegg1989phase}.\footnote{For small $\langle \hat{n} \rangle$, one may instead employ the canonical pair $\hat{n}$ and $\sin\hat{\Phi}$~\cite{vaccaro1990physical,pegg1989phase}.}

In direct analogy with $\hat{a}(s)$, the operator family
\begin{equation}
	\hat{b}_n(s)=\frac{\hat{n}+ i\, s\,  \gamma_0 \hat{\Phi}}{\sqrt{2 \gamma_0}}
	\label{bns_operator}
\end{equation}
has eigenstates that are number-squeezed for $s<1$ and phase-squeezed for $s>1$. These states are minimum-uncertainty (intelligent) number–phase states~\cite{vaccaro1990physical}.\footnote{Such constructions can be generalized to other canonical conjugate pairs.}

For $\langle \hat{n} \rangle \gg 1$, the eigenstates
\begin{equation}
	\hat{b}_n \,|\beta_n\rangle = \beta_n \,|\beta_n\rangle
	\label{eigenbn}
\end{equation}
of the operator $\hat{b}_n$ reduce to the usual coherent states~\cite{barnett1989hermitian,vaccaro1990physical,pegg1989phase,barnett2007quantum}. In this large–photon-number limit, one has $|\beta_n\rangle \rightarrow |\beta\rangle$, where the coherent-state amplitude $\beta$ is related to the eigenvalue $\beta_n$ via
\begin{equation}
	\beta = \sqrt{\mathrm{Re}{\beta_n}} \: \exp\big(\mathrm{Im}{\beta_n}\big),
\end{equation}
as shown in Refs.~\cite{vaccaro1990physical,tasgin2019anatomy,pegg1989phase,barnett1989hermitian,barnett2007quantum}.

Throughout this work, we operate in the regime $\langle \hat{n} \rangle \gg 1$, where the phase operator $\hat{\Phi}$ is well behaved and satisfies the canonical commutation relation $[\hat{n},\hat{\Phi}] = i$~\cite{vaccaro1990physical,tasgin2019anatomy,pegg1989phase,barnett1989hermitian,barnett2007quantum}.

%

For the numerical calculations, we define the phase operator $\hat{\Phi}$ using the Pegg–Barnett formalism~\cite{vaccaro1990physical,pegg1989phase,barnett1989hermitian,barnett2007quantum}. We explicitly verify the validity of the commutation relation and the coherent-state approximation. Even for moderate occupation numbers, $\langle \hat{n} \rangle \sim 30$, the commutator $[\hat{n},\hat{\Phi}] = i$ is satisfied to 3–4 significant digits, and the eigenstates $|\beta_n\rangle$ exhibit a high overlap with the corresponding coherent states~\cite{barnett1989hermitian,barnett2007quantum}.

Throughout the paper, $\hat{a}$ denotes the incident mode through which randomly displaced quadrature-squeezed states $|\alpha_i,r_i\rangle$ are injected. The quadrature squeezing in the $\hat{a}$ mode is converted into number squeezing in the $\hat{b}_n$ mode. We subsequently measure the $g^{(2)}(\tau)$ histogram of the $\hat{b}_n$ mode after the required number-manipulation steps (Fig.~\ref{fig1}).

\section{Analogous Displacement Operator} \label{sec:AnalogDisp}

Next, we introduce the analogous displacement operator
\begin{equation}
	\hat{D}_n(\beta_n)= \exp\big( \beta_n^* \hat{b}_n - \beta_n \hat{b}_n^\dagger \big).
\end{equation}
Using the Baker–Campbell–Hausdorff (Hadamard) lemma~\cite{Scully_Zubairy_1997} together with the commutation relation $[\hat{b}_n,\hat{b}_n^\dagger]=1$, one finds that $\hat{D}_n(\beta_n)$ acts as a displacement on $\hat{b}_n$:
\begin{equation}
	\hat{D}_n^\dagger(\beta_n)\, \hat{b}_n \,\hat{D}_n(\beta_n)
	= \hat{b}_n + \beta_n.
	\label{bn_betan}
\end{equation}

Thus, $\hat{D}_n(\beta_n)$ generates the transformations
\begin{equation}
	\hat{n} \; \rightarrow \; \hat{n} + \mathrm{Re}\{\beta_n\},
	\qquad
	\hat{\Phi} \; \rightarrow \; \hat{\Phi} + \mathrm{Im}\{\beta_n\},
\end{equation}
corresponding to displacements in the number–phase plane. This construction is directly analogous to the usual displacement operator $\hat{D}(\alpha)=\exp(\alpha \hat{a}^\dagger - \alpha^* \hat{a})$, which generates translations in the $x$–$p$ phase space. We also note that $\hat{D}_n(\beta_n)$ is unitary, $\hat{D}_n(\beta_n) \, \hat{D}_n^\dagger(\beta_n) = \hat{D}_n^\dagger(\beta_n) \, \hat{D}_n(\beta_n) = \hat{I}$, where $\hat{I}$ denotes the identity operator.


Below, using the above transformation, we show that $\hat{D}_n(\beta_n)$ acts on ordinary coherent states as a displacement,
\begin{equation}
	\beta = \sqrt{\mathrm{Re}\{\beta_n\}} \, \exp\big(\mathrm{Im}\{\beta_n\}\big),
\end{equation}
such that
\begin{equation}
	\hat{D}_n(\beta_n) \, |\alpha\rangle = |\alpha + \beta\rangle .
\end{equation}
This result will be used in Sec.~\ref{sec:noNc} to demonstrate that the $\hat{B}_n(\theta)$ operation cannot generate nonclassicality unless the input $\hat{a}$ mode is initially squeezed.

The eigenvalue equation~(\ref{eigenbn}) can be rewritten as
\begin{equation}
	\hat{D}_n(\beta_n) \, \hat{b}_n \, \hat{D}_n^\dagger(\beta_n) \,
	\hat{D}_n(\beta_n) \, |\alpha_n\rangle
	= \alpha_n \,\hat{D}_n(\beta_n) \, |\alpha_n\rangle ,
	\label{x-equation}
\end{equation}
where we define the displaced state
\begin{equation}
	|x\rangle \equiv \hat{D}_n(\beta_n) \, |\alpha_n\rangle .
\end{equation}
Using Eq.~(\ref{bn_betan}),
\begin{equation}
	\hat{D}_n(\beta_n) \,\hat{b}_n \,\hat{D}_n^\dagger(\beta_n)
	= \hat{b}_n - \beta_n ,
\end{equation}
Eq.~(\ref{x-equation}) becomes
\begin{equation}
	(\hat{b}_n - \beta_n) \,|x\rangle = \alpha_n \,|x\rangle .
\end{equation}
This may be rewritten as the eigenvalue equation
\begin{equation}
	\hat{b}_n \, |x\rangle = (\alpha_n + \beta_n) \, |x\rangle ,
\end{equation}
showing that the displacement operator $\hat{D}_n(\beta_n)$ shifts the eigenvalue of $\hat{b}_n$ by $\beta_n$.


Putting these results together, in the large-occupation regime $\langle \hat{n} \rangle \gg 1$ the eigenstates of $\hat{b}_n$~\footnote{In fact, the eigenstates of $\hat{b}_n$ are also classical states, e.g., when the $\hat{n}$ and $\sin(\hat{\Phi})$ operators are used as a conjugate pair. These states satisfy equal uncertainties and achieve the minimum uncertainty product. However, the transformation in Eq.~(\ref{bn_betan}) is valid only in the large–photon-number regime, $\langle \hat{n}\rangle \gg 1$, where the commutation relation $[\hat{n},\hat{\Phi}]=i$ holds.} coincide with the usual coherent states~\cite{barnett1989hermitian,vaccaro1990physical,tasgin2019anatomy,pegg1989phase,barnett2007quantum}. Prior to the application of the displacement operator $\hat{D}_n(\beta_n)$, the mode is in a coherent state $|\alpha_n\rangle \equiv |\alpha\rangle$, with complex amplitude
\begin{equation}
	\alpha = \sqrt{{\rm Re}\{\alpha_n\}} \,\exp \left({\rm Im}\{\alpha_n\}\right).
\end{equation}
After applying $\hat{D}_n(\beta_n)$, the state is mapped to the displaced eigenstate
\begin{equation}
	|x\rangle = |\alpha_n + \beta_n\rangle .
\end{equation}
Thus, $\hat{D}_n(\beta_n)$ transforms an initial coherent state $|\alpha\rangle$ into another coherent state
\begin{equation}
	|\alpha'\rangle \equiv |\alpha_n + \beta_n\rangle ,
\end{equation}
provided that the displaced state remains within the large-$\langle\hat{n}\rangle$ regime. The corresponding coherent-state amplitude is
\begin{equation}
	\alpha' = \sqrt{{\rm Re}\{\alpha_n + \beta_n\}} \,
	\exp\left({\rm Im}\{\alpha_n + \beta_n\}\right).
\end{equation}

This result and the methods we use here  will be essential in Sec.~\ref{sec:noNc}, where we show that the $\hat{B}_n(\theta)$ operation alone cannot generate nonclassicality.

\section{Analogous Beam-Splitter Operator} \label{sec:BSn}

The standard beam-splitter (BS) operator
\begin{equation}
	\hat{B}(\theta)=\exp \left[\theta \big( e^{i\phi} \hat{a}^\dagger \hat{b}
	- e^{-i\phi} \hat{b}^\dagger \hat{a} \big) \right]
	\label{Boperator}
\end{equation}
is well known to transfer quadrature squeezing from one optical mode to another, where it remains quadrature squeezing~\cite{ge2015conservation}. Closely related BS-like interactions also enable the mapping of optical quadrature squeezing onto collective spin degrees of freedom, producing spin squeezing in atomic ensembles~\cite{sorensen2001many,kuzmich1997spin,tacsgin2011spin}. A representative example is the interaction Hamiltonian
\begin{equation}
	\hat{\mathcal{H}}_{\rm spin}
	= \hbar g \big( \hat{S}_- \hat{a} + \hat{a}^\dagger \hat{S}_+ \big),
\end{equation}
where $\hat{S}_{\pm}$ are the collective spin raising and lowering operators~\cite{sorensen2001many,kuzmich1997spin,tacsgin2011spin}.

Here, we introduce the analogous beam-splitter transformation
\begin{equation}
	\hat{B}_n(\theta)
	= \exp\left[ \theta \big( e^{i\phi} \hat{b}_n^\dagger \hat{a}
	- e^{-i\phi} \hat{a}^\dagger \hat{b}_n \big) \right],
	\label{Bn_operator}
\end{equation}
which couples the optical annihilation operator
$\hat{a} = (\hat{x} + i\hat{p})/\sqrt{2}$ to $ \hat{b}_n=( \hat{n}+ i \gamma_0 \hat{\Phi} )/\sqrt{2\gamma_0}$,
associated with the canonically conjugate number–phase variables. This transformation transfers squeezing from the quadrature degrees of freedom of mode $\hat{a}$ into the number–phase sector of mode $\hat{b}_n$.

In this work, we specifically demonstrate the conversion of quadrature squeezing in $\hat{a}$ into {\it number squeezing} in $\hat{b}_n$, and quantify the resulting nonclassicality via the second-order correlation function $g^{(2)}(0)$ of the $\hat{b}_n$ mode.

%

Using the Baker–Campbell–Hausdorff formula~\cite{Scully_Zubairy_1997} and the commutation relation
\(
[\hat{b}_n,\hat{b}_n^\dagger]=1,
\)
the operator \(\hat{B}_n(\theta)\) transforms the annihilation operators as
\begin{eqnarray}
	\hat{a}(\xi)
	&=&
	\hat{B}_n(\theta)\,\hat{a}\,\hat{B}_n^\dagger(\theta)
	=
	t\,\hat{a}+r\,\hat{b}_n,
	\label{axi}
	\\
	\hat{b}_n(\xi)
	&=&
	\hat{B}_n(\theta)\,\hat{b}_n\,\hat{B}_n^\dagger(\theta)
	=
	t\,\hat{b}_n-r\,\hat{a},
	\label{bnxi}
\end{eqnarray}
where \(t=\cos\theta\) and \(r=\sin\theta\).
In Sec.~\ref{sec:SqzTransfer}, we show that the transformation \(\hat{B}_n(\theta)\)
converts initial quadrature squeezing—generated by the usual squeezing Hamiltonian
\(\mathcal{H}\propto (\hat{a}^2+\text{H.c.})\)—into number squeezing.
The latter is associated with an effective Hamiltonian
\(\mathcal{H}_n\propto (\hat{b}_n^2+\text{H.c.})\), or equivalently with the squeezing operator
\begin{equation}
	\hat{S}_n(\xi)
	=
	\exp\!\left[
	\frac{1}{2}\left(
	\xi\,\hat{b}_n^2-\xi^*\,\hat{b}_n^{\dagger 2}
	\right)
	\right],
	\label{Snxi}
\end{equation}
that will be introduced in Sec.~\ref{sec:AnalSqz}.

In Sec.~\ref{sec:SqzTransfer}, we analytically demonstrate how the action of
\(\hat{B}_n(\theta)\) continuously maps the standard quadrature-squeezing transformation
$\hat{S}(z) = \exp[(z^*\hat{a}^2 - z\hat{a}^{\dagger 2})/2]$ onto the number–phase squeezing
transformation \(\hat{S}_n(\xi)\), defined in Eq.~(\ref{Snxi}).
Our numerical simulations further confirm that, under the action of \(\hat{B}_n(\theta)\),
quadrature squeezing in the \(\hat{a}\) mode decreases while number squeezing in the
\(\hat{b}_n\) mode increases (see Fig.~\ref{fig2}).


Before proceeding, we must show that the transformation $\hat{B}_n(\theta)$—analogously to an ordinary beam splitter~\cite{kim2002entanglement,asboth2005computable,adesso2010quantum,ge2015conservation}—cannot generate nonclassicality by itself. This property is essential for our protocol: it ensures that any observed nonclassical signature, such as $g^{(2)}<1$, must originate from pre-existing quadrature squeezing in the incoming states rather than from the conversion process itself.

\section{$\mathbf{\hat{B}_n(\theta)}$ does not generate nonclassicality} \label{sec:noNc}

In this section, we show that $\hat{B}_n(\theta)$ does not create nonclassicality, but merely converts one form of nonclassicality into another. Specifically, it transforms quadrature squeezing in the $\hat{a}$ mode into number/phase squeezing in the $\hat{b}_n$ mode. If the $\hat{a}$ mode is not initially squeezed, $\hat{B}_n(\theta)$ cannot generate nonclassical features. This leads to an important conclusion: observing $g^{(2)}(0)<1$ in a histogram measurement implies that the received copies must originate from a quadrature-squeezed state.


Our demonstration closely follows the reasoning of Ref.~\cite{kim2002entanglement}. We assume that the initial two-mode state is classical. Its density matrix can therefore be written as~\cite{duan2000inseparability,brunelli2015single}
\begin{equation}
	\hat{\rho}_{\rm c} = \sum_k P_k \, |\alpha^{(k)}\rangle |\beta_n^{(k)}\rangle
	\langle\alpha^{(k)}| \langle \beta_n^{(k)} |,
	\label{dens_class}
\end{equation}
where the first mode is in a coherent state $|\alpha^{(k)}\rangle$ with classical probability $P_k$.  The second mode is likewise in the coherent state $|\beta_n^{(k)}\rangle$  with the same probability $P_k$.

We restrict ourselves to the large-occupation regime, where $|\beta_n^{(k)}\rangle$ is effectively equivalent to an ordinary coherent state $|\beta^{(k)}\rangle$~\cite{barnett1989hermitian,vaccaro1990physical,tasgin2019anatomy,pegg1989phase,barnett2007quantum}.
The corresponding complex amplitude is given by~\cite{vaccaro1990physical,barnett2007quantum}
\begin{equation}
	\beta^{(k)}= \sqrt{ {\rm Re} \{ \beta_n^{(k)} \} } \times \exp \left({\rm Im}\{ \beta_n^{(k)} \}  \right).
\end{equation}
For such states, the Glauber–Sudarshan $P$ function is positive everywhere, and the state is therefore classical~\cite{kim2002entanglement,asboth2005computable,adesso2010quantum}.

We aim to show that the density matrix transformed by $\hat{B}_n(\theta)$,
\begin{equation}
	\hat{\rho} = \hat{B}_n(\theta) \, \hat{\rho}_{\rm c} \, \hat{B}_n^\dagger(\theta),
	\label{dens_transformed}
\end{equation}
remains classical. This can be shown in two equivalent ways.

($\mathtt{1}$)--- The coherent states appearing in Eq.~(\ref{dens_class}) are simultaneous eigenstates of $\hat{a}$ and $\hat{b}_n$ in the large-occupation limit $\langle\hat{n}\rangle \gg 1$. Explicitly,
\begin{eqnarray}
	\hat{a} \, |\alpha\rangle |\beta_n\rangle
	&=& \alpha \, |\alpha\rangle |\beta_n\rangle,
	\label{ap}
	\\
	\hat{b}_n \, |\alpha\rangle |\beta_n\rangle
	&=& \beta_n \, |\alpha\rangle |\beta_n\rangle .
	\label{bnp}
\end{eqnarray}

We now apply the operator $\hat{B}_n(\theta)$ to both equations. Inserting the identity $\hat{I} = \hat{B}_n^\dagger(\theta) \hat{B}_n(\theta)$ on the left-hand side yields
\begin{eqnarray}
	\left( \hat{B}_n(\theta) \, \hat{a} \, \hat{B}_n^\dagger(\theta) \right)
	\hat{B}_n(\theta) \, |\alpha\rangle |\beta_n\rangle
	&=& \alpha \, \hat{B}_n(\theta) \, |\alpha\rangle |\beta_n\rangle , \quad
	\label{BnEigena}
	\\
	\left( \hat{B}_n(\theta) \, \hat{b}_n \, \hat{B}_n^\dagger(\theta) \right)
	\hat{B}_n(\theta) \, |\alpha\rangle |\beta_n\rangle
	&=& \beta_n \, \hat{B}_n(\theta) \, |\alpha\rangle |\beta_n\rangle . \quad
	\label{BnEigenbn}
\end{eqnarray}
Thus, the transformed states $|x\rangle \equiv \hat{B}_n(\theta) \,  |\alpha\rangle| \beta_n  \rangle$ remain eigenstates of the transformed field operators.
%
%
%
%
%
%
%
%
%
%
%
%
These are the eigenvalue equations for the transformed operators
\begin{eqnarray}
\hat{a}'(\theta)=\hat{B}_n(\theta) \, \hat{a} \, \hat{B}_n^\dagger(\theta)=\cos\theta \: \hat{a} + e^{i\phi} \sin\theta \: \hat{b}_n, \quad
\\
\hat{b}_n'(\theta)=\hat{B}_n(\theta) \, \hat{b}_n \, \hat{B}_n^\dagger(\theta) =\cos\theta \: \hat{b}_n - e^{-i\phi} \sin\theta \: \hat{a}.  \quad
\end{eqnarray}
They share a common eigenstate 
\begin{equation} 
	|x\rangle = \hat{B}_n(\theta)\, |\alpha\rangle |\beta_n\rangle . 
\end{equation} One can verify that Eqs.~(\ref{BnEigena}) and (\ref{BnEigenbn}) are satisfied by the coherent product state 
\begin{equation} 
	|x\rangle = |\alpha'\rangle |\beta_n'\rangle , 
\end{equation} with 
\begin{eqnarray} 
	\alpha' &=& \cos\theta \, \alpha - e^{i\phi} \sin\theta \, \beta_n , \label{alphap} \\ 
	\beta_n' &=& \cos\theta \, \beta_n + e^{-i\phi} \sin\theta \, \alpha . \label{betanp} \end{eqnarray} 
Thus, the beam-splitter–transformed state 
\begin{equation} 
	|x\rangle \equiv \hat{B}_n(\theta)\, |\alpha\rangle |\beta_n\rangle , 
\end{equation} which is the common eigenstate of \(\hat{a}'(\theta)\) and \(\hat{b}_n'(\theta)\), is again a coherent state.

Including the $k$ index, that is, the states  $|\alpha^{(k)}\rangle|\beta_n^{(k)}\rangle$, and substituting them into Eqs.~(\ref{dens_class}) and (\ref{dens_transformed}), we conclude that the $\hat{B}_n(\theta)$-transformed density matrix $\hat{\rho}$ in Eq.~(\ref{dens_transformed}) is also classical. Therefore, if the incident copies in the $\hat{a}$ mode do not contain any nonclassicality, the operation $\hat{B}_n(\theta)$ cannot generate nonclassicality.

%

($\mathtt{2}$)---  Alternatively, one can reach the same conclusion by noting that the initial coherent product state
\begin{equation}
	|\psi\rangle=| \alpha^{(k)} \rangle | \beta_n^{(k)} \rangle
\end{equation}
is also a common eigenstate of the transformed (time-evolved) operators $\hat{a}'(\theta)$ and $\hat{b}_n'(\theta)$ obtained via Eqs.~(\ref{ap}) and (\ref{bnp}). As a result, any normally ordered correlation functions satisfy
\begin{eqnarray}
\langle x | (\hat{a}^\dagger{})^m \, (\hat{a})^\ell  |x \rangle = \langle \alpha | \langle\beta_n| (\hat{a}'^\dagger{})^m \, (\hat{a}')^\ell  |\alpha \rangle |\beta_n\rangle ,
\\
\langle x | (\hat{b}_n^\dagger{})^m \, (\hat{b}_n)^\ell  |x \rangle = \langle \alpha | \langle\beta_n| (\hat{b}_n'^\dagger{})^m \, (\hat{b}'_n)^\ell  |\alpha \rangle |\beta_n\rangle ,
\end{eqnarray}
and they factorize exactly as they do for a classical field:
\begin{equation}
	(\alpha'^*)^m (\alpha')^\ell  \quad \text{and} \quad (\beta_n'^*)^m \, (\beta_n)^\ell.
\end{equation}
Such factorizations are a unique property of coherent states~\cite{MandelWolf,Louisell1973,glauber1963coherent} and confirm that the field remains classical after the $\hat{B}_n(\theta)$ transformation.

%
%

Finally, we emphasize once more that in our derivations—and in the setup we consider (Fig.~\ref{fig1})—we assume that the $\hat{B}_n(\theta)$ operation keeps the $\hat{b}_n$-mode in the large-$\langle\hat{n}\rangle$ regime. This ensures that the commutation relation $[\hat{n},\hat{\Phi}] = i$ holds and that the transformations in Eqs.~(\ref{axi}) and (\ref{bnxi}) remain valid.


($\mathtt{3}$)--- As a third alternative, one can directly adopt the displacement plus beam-splitter operator approach demonstrated in Ref.~\cite{kim2002entanglement}.

\section{Analytical Demonstration of the Squeezing Transfer} \label{sec:SqzTransfer}

%

In the previous section, we showed that $\hat{B}_n(\theta)$ cannot generate nonclassicality from classical states. In this section, we analytically show how $\hat{B}_n(\theta)$ transfers the initial quadrature squeezing into number–phase squeezing in the second mode. We consider an initially quadrature-squeezed state in the $\hat{a}$ mode and explicitly show how $\hat{B}_n(\theta)$ transforms it into a number-squeezing operator.

We take the first mode to be in a squeezed coherent state  $|z,\alpha\rangle = \hat{S}(z) \,|\alpha\rangle$, and the second mode to be in a strongly occupied coherent state $|\beta_n\rangle$. The initial two-mode state is therefore
\begin{equation}
	|\psi_0\rangle = |z,\alpha\rangle \,|\beta_n\rangle,
\end{equation}
where $\hat{S}(z) = \exp[(z^*\hat{a}^2 - z\hat{a}^{\dagger 2})/2]$ is the single-mode squeezing operator~\cite{Scully_Zubairy_1997}, with $z = \mathtt{r} \, e^{i\varphi}$ is
the squeezing parameter~\cite{ScullyZubairyBook}. We now study how this state transforms under  $\hat{B}_n(\theta)$,
\begin{equation}
	|\psi(\theta)\rangle
	= \hat{B}_n(\theta) \, |\mathtt{r},\alpha\rangle |\beta_n\rangle
	= \hat{B}_n(\theta) \, \hat{S}(\mathtt{r}) \, |\alpha\rangle |\beta_n\rangle .
\end{equation}

By inserting the identity operator $\hat{I} = \hat{B}_n^\dagger(\theta) \hat{B}_n(\theta)$, the transformed state can be written as
\begin{equation}
	|\psi(\theta)\rangle
	= \bigl(\hat{B}_n(\theta) \, \hat{S}(\mathtt{r}) \, \hat{B}_n^\dagger(\theta)\bigr)
	\, \hat{B}_n(\theta) \, |\alpha\rangle |\beta_n\rangle .
\end{equation}
From the previous section, we know that $\hat{B}_n(\theta) \, |\alpha\rangle|\beta_n\rangle$ is a coherent product state $|\alpha'\rangle|\beta_n'\rangle$ where $\alpha'$ and $\beta_n'$ are given in Eqs.~(\ref{alphap}) and (\ref{betanp}). The remaining task is therefore to determine how the operator
\begin{equation}
	\hat{B}_n(\theta) \, \hat{S}(\mathtt{r}) \, \hat{B}_n^\dagger(\theta)
\end{equation}
acts on the coherent state $|\alpha'\rangle|\beta_n'\rangle$.

For simplicity, we ignore the phase factors $e^{i\phi}$ and $e^{i\varphi}$. The transformed squeezing operator then takes the form
\begin{eqnarray}
	\hat{B}_n(\theta) \hat{S}(r) \hat{B}_n^\dagger(\theta) = \exp \Big\{  \frac{1}{2} \mathtt{r} \big[ &&  t^2(\hat{a}^2-\hat{a}^\dagger{}^2 ) + r^2(\hat{b}_n^2-\hat{b}_n^\dagger{}^2) \nonumber
	\\
	&& + 2tr(\hat{a}\hat{b}_n-\hat{b}_n^\dagger \hat{a}^\dagger) \big] \Big\},
	\label{squeezing_transfer}
\end{eqnarray}
where $t=\cos\theta$, $r=\sin\theta$ and $\mathtt{r}$ is the initial quadrature squeezing strength.

The first term on the right-hand side of Eq.~(\ref{squeezing_transfer}) is the usual quadrature-squeezing operator~\cite{Scully_Zubairy_1997}. The second term is the analogous number–phase squeezing operator,
\begin{equation}
	\hat{S}_n(\xi)
	= \exp\left((\xi \hat{b}_n^2 - \xi^* \hat{b}_n^{\dagger 2})/2\right),
\end{equation}
whose squeezing properties are analyzed in the next section. As $\hat{B}_n(\theta)$ acts for a longer time (that is, as $\theta$ increases), the coefficient of the quadrature-squeezing term $\exp[{\tt r} \, t^2(\hat{a}^2-\hat{a}^\dagger{}^2) ]$ decreases, while the strength of the number–phase squeezing term $\exp[ {\tt r}\, r^2(\hat{b}_n^2-\hat{b}_n^\dagger{}^2)]$ increases. Since $\hat{B}_n(\theta) \, \hat{S}(r) \hat{B}_n^\dagger(\theta)$ acts on the coherent state $|\alpha'\rangle|\beta_n'\rangle$, this shows directly that the initial quadrature squeezing in mode $\hat{a}$ is gradually converted into number–phase squeezing in mode $\hat{b}_n$ as $\theta$ grows. This behavior is already observed in our numerical simulations, as shown in Fig.~\ref{fig2}.

The third term on the right-hand side of Eq.~(\ref{squeezing_transfer}) is analogous to a two-mode squeezing interaction. In the standard case, the operator $\exp[ z(\hat{a}^\dagger\hat{b}-\hat{b}^\dagger \hat{a} ) ]$  creates entanglement between the quadratures of two modes. Here, however, $\exp \{ {\tt r} \, 2tr(\hat{a}\hat{b}_n-\hat{b}_n^\dagger \hat{a}^\dagger) \}$ generates entanglement between the quadrature of the first mode and the number–phase observables of the second mode. In other words, a measurement of the relevant quadrature of the $\hat{a}$ mode has its main effect on the number variable of the $\hat{b}_n$ mode.

In the standard setting, entanglement is witnessed through the squeezing of collective operators such as $\hat{u}=\hat{x}_1+c\, \hat{x}_2$ and $\hat{v}=\hat{p}_1-c\, \hat{p}_2$
which satisfy the Duan inseparability criterion~\cite{duan2000inseparability,tasgin2019anatomy}, where $c$ is a real number. In the present case, since the entanglement involves the quadrature of the first mode and the number–phase variables of the second mode, an analogous inseparability criterion is expected to involve collective operators of the form $\hat{u}_n$=$\hat{x}_1 + c \: \hat{n}_b$ where $\hat{n}_b$ is the number operator of the second mode. A derivation following the same steps as in Ref.~\cite{duan2000inseparability} can therefore be carried out for such $\hat{u}_n$-type operators.

\section{Analogous  Squeezing Operator} \label{sec:AnalSqz}

In this section, we introduce an analogous squeezing operator $\hat{S}_n(\xi)$ that acts in the number–phase plane. We show that this operator reduces the uncertainty in the photon number at the expense of an increased phase uncertainty, or vice versa.

We define the number/phase–squeezing operator as
\begin{equation}
	\hat{S}_n(\xi)= \exp\left[\frac{1}{2}\left(\xi \,\hat{b}_n^{2}-\xi^{*} \,\hat{b}_n^{\dagger 2}\right)\right],
	\label{Sn}
\end{equation}
in direct analogy with the standard quadrature–squeezing operator $\hat{S}(z)=\exp\left( z(\hat{a}^2-\hat{a}^\dagger{}^2)/2\right)$~\cite{Scully_Zubairy_1997}.
The operator $\hat{S}_n(\xi)$ transforms the mode operator $\hat{b}_n$ according to
\begin{equation}
	\hat{b}_n(\xi)
	= \hat{S}_n^\dagger(\xi) \,\hat{b}_n \,\hat{S}_n(\xi)
	= \cosh{\tt r} \, \hat{b}_n
	- e^{i\varphi'} \sinh{\tt r} \, \hat{b}_n^\dagger,
	\label{bnxi2}
\end{equation}
where $\xi=\mathtt{r}\, e^{i\varphi'}$. This result follows directly from the Baker–Campbell–Hausdorff formula~\cite{Scully_Zubairy_1997}, using the large–$\langle \hat{n}\rangle$ commutation relation $[\hat{n},\hat{\Phi}]=i$.

Using Eq.~\eqref{bnxi2}, one can determine the transformation of the number and phase operators. For $\varphi'=0$, they transform as
\begin{equation}
	\hat{n}(\xi) = e^{-{\tt r}} \,\hat{n},
	\qquad
	\hat{\Phi}(\xi) = e^{+{\tt r}} \, \hat{\Phi},
	\label{n-squeezing}
\end{equation}
demonstrating that the photon-number uncertainty is squeezed while the phase uncertainty is correspondingly increased. For $\varphi'=\pi/2$, the opposite behavior occurs.

More generally, the operator $\hat{S}_n(\xi)$ generates squeezing in the $n$-$\Phi$ plane along an axis determined by the phase $\varphi'$, in full analogy with conventional quadrature squeezing~\cite{Scully_Zubairy_1997}.

Together, Eq.~\eqref{n-squeezing} and the transformation given in Eq.~\eqref{squeezing_transfer} explicitly demonstrate that the initial quadrature squeezing of the first mode is converted into photon-number (or phase) squeezing in the second mode.

\section{Numerical Simulations} \label{sec:numeric}

As discussed in the previous sections, we also perform numerical simulations to demonstrate the transformation of quadrature squeezing in the first mode into photon-number squeezing in the second mode using a two-dimensional, finite-size Fock space. We implement the operator $\hat{B}_n(\theta)$ in a two-mode Hilbert space with maximum Fock-state dimensions $n_1^{(\rm max)}=100$ and $n_2^{(\rm max)}=100$. This corresponds to sparse matrices of dimension
\begin{equation}
	n_1^{(\mathrm{max})} n_2^{(\mathrm{max})} \times n_1^{(\mathrm{max})} n_2^{(\mathrm{max})}
	= 10{,}000 \times 10{,}000 .
\end{equation}
For the phase operator $\hat{\Phi}$, we employ the Pegg–Barnett formalism~\cite{vaccaro1990physical,pegg1989phase,barnett2007quantum}.

\begin{figure}
	\centering
	\includegraphics[width=0.48 \textwidth]{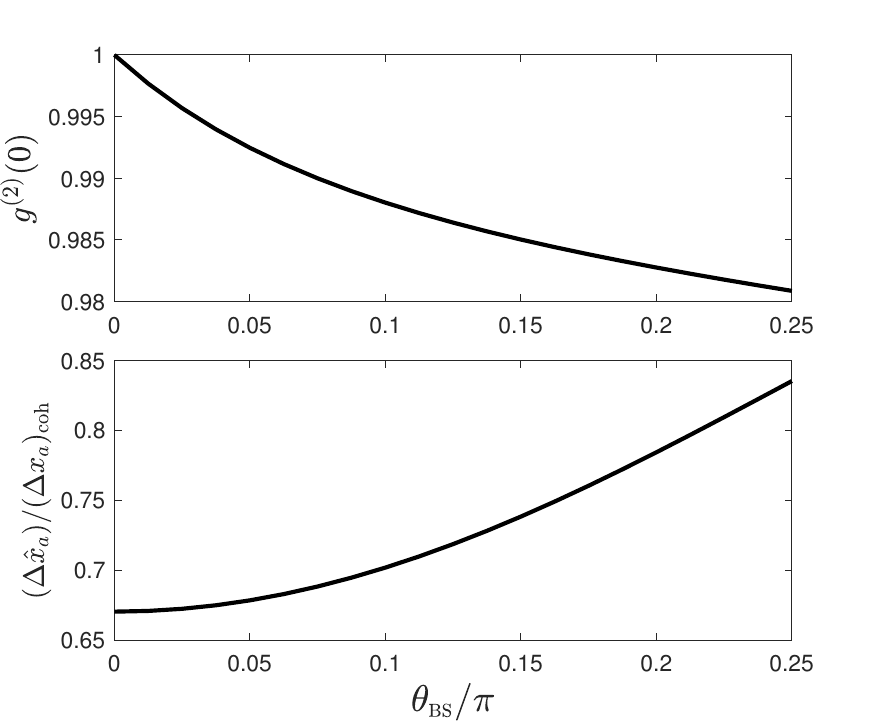}
	\caption{Transformation of initial quadrature squeezing into number-squeezing ($g^{(2)}(0)<1$) under the action of the $\hat{B}_n(\theta)$ operator. (a) Nonclassicality of the second ($\hat{b}_n$) mode increases as $g^{(2)}(0)$ decreases. (b) Quadrature squeezing of the first ($\hat{a}$) mode decreases, with the uncertainty $(\Delta \hat{x}_a)$ increasing. One type of nonclassicality (quadrature squeezing) present in the first mode is converted into another type (number-squeezing) in the second mode.}
	\label{fig2}
\end{figure}

In Fig.~\ref{fig2}, we present a numerical illustration of the squeezing-transfer mechanism analytically studied in Eqs.~(\ref{squeezing_transfer}) and~(\ref{n-squeezing}). The initial quadrature squeezing of the first mode is progressively transformed into photon-number squeezing in the second mode under the action of the generalized beam-splitter operator $\hat{B}_n(\theta)$. As $\theta$ increases, the quadrature squeezing in the first mode is gradually consumed (Fig.~\ref{fig2}b), while the nonclassicality transferred to the second mode manifests itself as number squeezing (Fig.~\ref{fig2}a).

More specifically, the initial squeezing of the quadrature operator $\hat{x}=(\hat{x}+i\,\hat{p})/\sqrt{2}$ diminishes, as indicated by the increase of its uncertainty, whereas the nonclassicality of the $\hat{b}_n$ mode increases, signaled by the second-order correlation function $g^{(2)}(0)$ decreasing below unity.

This behavior is reminiscent of the transfer of nonclassicality through an ordinary beam splitter~\cite{ge2015conservation,tasgin2020quantifications,liu2024classical}. The crucial difference is that the operator $\hat{B}_n(\theta)$ not only transfers nonclassicality between the modes, but also converts quadrature squeezing in the first mode into photon-number squeezing in the second mode. In contrast, a conventional beam splitter can transfer quadrature squeezing between modes but cannot convert quadrature squeezing into number squeezing.

We note that our numerical calculations are subject to a limited region of validity. During the squeezing-transfer process, the occupation $\langle \hat{n}\rangle$
of the second mode—initially prepared in a well-occupied state—decreases, as also predicted analytically in Eqs.~(\ref{squeezing_transfer}) and~(\ref{n-squeezing}). For sufficiently large transferred squeezing, $\langle \hat{n}\rangle$ eventually enters a low-occupation regime in which the approximate commutation relations employed in our analysis are no longer valid.

In our simulations, we therefore restrict the dynamics to a parameter range in which both modes remain well occupied. This behavior is explicitly monitored numerically, and for this reason the action of the operator $\hat{B}_n(\theta)$ in Fig.~\ref{fig2} is limited to a restricted range of $\theta$.

Finally, we have numerically verified that no photon-number squeezing is generated when the $\hat{a}$ mode is initialized in a classical (coherent) state, confirming that the observed effect arises from genuinely nonclassical input squeezing.

%

\begin{figure}
	\centering
	\includegraphics[width=0.48 \textwidth]{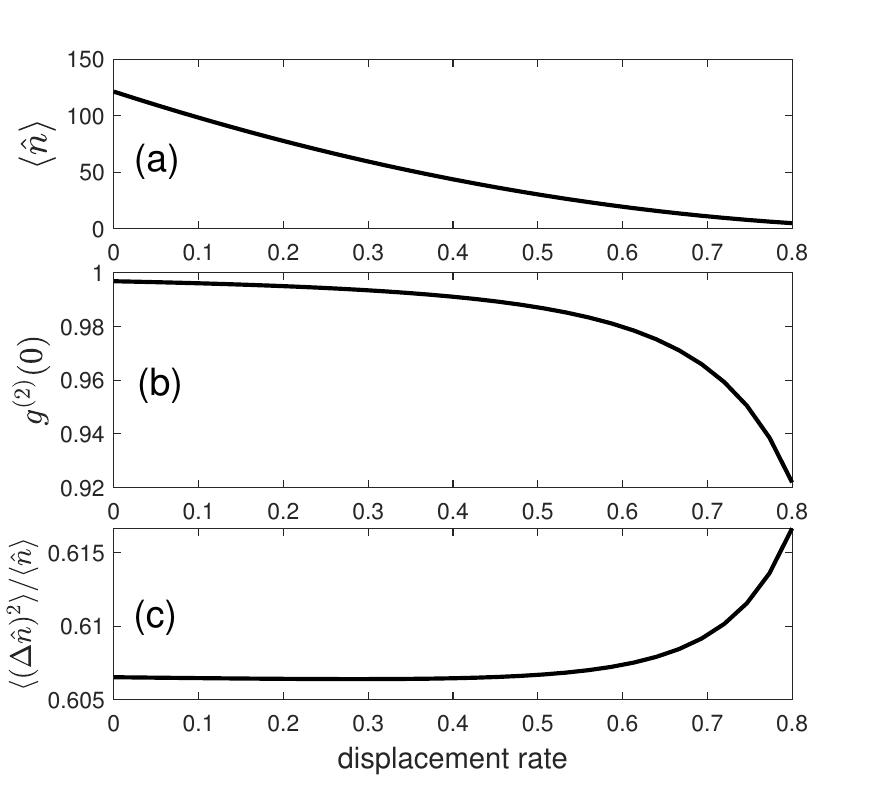}
	\caption{While the number-squeezed $\hat{b}_n$ output mode possesses a  $\langle(\Delta\hat{n})^2\rangle / \langle \hat{n}\rangle $ ratio much below the unity, the $g^{(2)}(0)$ stays close to unity because of the large occupation. We perform a counter-displacement in the $\hat{b}_n$ output mode before we send them to the $g^{(2)}(\tau)$ histogram. The $\hat{D}(\alpha')$ displacement does not generate nonclassicality from coherent state $\hat{b}_n$ outputs, however it makes $g^{(2)}(0)<1$ states more visible.
	}
	\label{fig3}
\end{figure}

\section{Making $g^{(2)}(0)$ more visible} \label{sec:counterDisplacement}

As can be seen in Fig.~\ref{fig2}, the value of $g^{(2)}(0)$ remains close to unity. This does not imply weak number squeezing. In fact, the system exhibits strong photon-number squeezing: the normalized number variance $\langle(\Delta\hat{n})^2\rangle / \langle \hat{n}\rangle$
lies well below unity, indicating sub-Poissonian statistics. However, because of the relation between the second-order correlation function and the number variance $F=\langle(\Delta\hat{n})^2\rangle / \langle \hat{n}\rangle$~\cite{Scully_Zubairy_1997,MandelWolf},
\begin{equation}
	g^{(2)}(0) = 1 + \frac{F-1}{\langle \hat{n}\rangle},
\end{equation}
the quantity $g^{(2)}(0)$ remains close to unity whenever the photon occupation $\langle \hat{n}\rangle \gg 1$, even in the presence of very strong number squeezing. In our scheme, the experimentally accessible quantity is $g^{(2)}(0)$, rather than the normalized number variance $F$ itself.

To overcome this limitation, we apply a counter-displacement $\hat{D}(\alpha')$ to the output mode $\hat{\rho}_b$ before constructing the photon-number histogram; see Fig.~\ref{fig1}. This operation reduces the photon occupation $\langle\hat{n}\rangle$ of the displaced state $\hat{\rho}_{b'}$, thereby amplifying the deviation of $g^{(2)}(0)$
from unity. As a result, $g^{(2)}(0)$ drops well below one, as shown in Fig.~\ref{fig3}. In particular, Fig.~\ref{fig3} demonstrates that even a value of $g^{(2)}(0)$ extremely close to unity (e.g., $g^{(2)}(0)=0.995$) can be converted into a clearly visible sub-Poissonian signal through counter-displacement. Consequently, the already smaller values (e.g., 0.98) obtained in Fig.~\ref{fig2}a can be made dramatically more pronounced.

We emphasize that the displacement operator does not generate nonclassicality~\cite{kim2002entanglement,simon1994quantum,asboth2005computable,adesso2010quantum,GAdesso04,serafini2003symplectic} unless nonclassical features are already present in the state $\hat{\rho}_b$. The operation  $\hat{D}(\alpha')$ neither destroys existing nonclassicality nor creates new nonclassicality. Rather, it renders the sub-Poissonian character  $g^{(2)}(0)<1$ more directly observable (Fig.~\ref{fig3}b), at the expense of increasing the absolute photon-number squeezing (Fig.~\ref{fig3}c).

Finally, we note that operating in the large-occupation regime is required only during the $\hat{B}_n(\theta)$ transformation. (This requirement arises because the theorem in Sec.~\ref{sec:noNc}, which relies on the commutation relation $[\hat{n},\hat{\Phi}]=i$, is valid only when $\langle\hat{n}\rangle$ is large where the phase operator is well defined.) However,  after the $\hat{B}_n(\theta)$ transformation—i.e., once quadrature squeezing has been converted into photon-number squeezing—the $\hat{b}_n$ mode no longer needs to remain in the high-occupation regime.

\section{$ \bf g^{(2)}(\mathbf{\tau})$ Histogram} \label{sec:histogram}

In this section, we first explain how a $g^{(2)}(\tau)$ histogram is obtained. We then present three methods to reliably resolve the nonclassical condition $g^{(2)}(0)<1$, even in the presence of pulse-to-pulse intensity fluctuations among the received copies, denoted by $\hat{\rho}_t^{(i)}$ in Fig.~\ref{fig1}. Two of these methods employ number-resolving photon detectors, while the third—an elegant method—uses standard single-photon detectors. In this third method, the relative pulse-to-pulse number fluctuations are reduced by several orders of magnitude without altering the value of $g^{(2)}(0)$.


\subsection{A $\mathbf{ g^{(2)}(\tau) }$  histogram}
A $g^{(2)}(\tau)$ histogram is measured using a Hanbury Brown–Twiss (HBT) interferometer. The incident single-mode field is split into two by a 50/50 beam splitter and detected by two detectors, $D_1$ and $D_2$. Correlations between the detection times are recorded. Detection events are grouped into time bins of width $\Delta\tau$, which is typically set by the detector dead time (on the order of nanoseconds).

The number of coincidence events within the time bin $[\tau,\tau+\Delta\tau]$ is denoted by ${\cal C}(\tau)$, where $\tau$ is the time delay between the two detector clicks. Coincidences are accumulated over a long integration time $T$, during which many pulses of the source contribute. The coincidence rate is given by
\begin{equation}
	R_{\rm coin}(\tau)= \frac{{\cal C}(\tau)}{T}.
\end{equation}
This rate is compared with the accidental coincidence rate,
\begin{equation}
	R_{\rm acc}= S_1 \, S_2 \, \Delta\tau,
\end{equation}
where $S_1$ and $S_2$ are the single-count rates at detectors $D_1$ and $D_2$. The second-order correlation function is then
\begin{equation}
	g^{(2)}(\tau) = \frac{R_{\rm coin}(\tau)}{R_{\rm acc}}
	= \frac{{\cal C}(\tau)}{S_1 \, S_2 \, \Delta\tau \, T}.
\end{equation}

\subsection{Copy-by-copy averaging}

If all detection events are aggregated before averaging, the measured quantity becomes
\begin{equation}
	g^{(2)}_{\rm agg}(0)
	= \frac{{\rm Avg} \left[\langle \hat{n}_i(\hat{n}_i-1)\rangle\right]}
	{\left({\rm Avg}\left[\langle \hat{n}_i\rangle\right]\right)^2},
\end{equation}
where $\hat{n}_i$ is the photon number in the $i$-th copy. In this case, pulse-to-pulse number fluctuations appear as classical intensity noise and reduce the visibility of the nonclassical condition $g^{(2)}(0)<1$.

A better approach is to evaluate coincidences on a copy-by-copy basis and then average the result,
\begin{equation}
	g^{(2)}(0)
	= {\rm Avg}\left[
	\frac{\text{coincidences in copy } i}{(I_i)^2}
	\right],
	\label{copy-by-copy}
\end{equation}
where $I_i$ s the total intensity of the $i$-th copy. This procedure removes the effect of pulse-to-pulse number fluctuations. We use this copy-by-copy approach in all three methods described below.

\subsection{Displacement and counter-displacement}

Although a displacement operation $\hat{D}(\alpha)$ already narrows down the relative pulse-to-pulse number fluctuations, after the $\hat{B}_n(\theta)$ operation we apply a counter-displacement $\hat{D}(\alpha')$ to the $\hat{b}_n$ output (transforming $\hat{\rho}_b$ into $\hat{\rho_{b'}}$) to make the nonclassical behavior in $g^{(2)}(0)$ more visible (see Fig.~\ref{fig3}b).

As an example, suppose that $\hat{D}(\alpha')$ reduces the mean photon number from $\langle \hat{n}_i( \hat{\rho}_{b} )\rangle=200$ to $\langle \hat{n}_i( \hat{\rho}_{b'} )\rangle=20$, while the number fluctuations introduced by the dynamical channel are$(\Delta \hat{n}_i) \sim 1$.
Although the relative fluctuations increase by a factor of 10, they remain small,  $(\Delta \hat{n}_i)/\langle\hat{n}\rangle=1/20$,
and are still much smaller than the original fluctuations of the incident copies.

\subsection{Our $\mathbf{g^{(2)}(0)}$ detection methods}
One can employ the following methods in our setup to accurately resolve the desired quantity $g^{(2)}(0)$.

{\it Method 1: Number-resolving detectors}.— In this method, as an example, we operate at a mean photon number $\langle \hat{n} \rangle \sim 20$, where number-resolving photon detectors (NRPDs) are capable of resolving photon numbers up to $\sim 20$~\cite{sempere2022reducing,low2025measurement}. Using Eq.~(\ref{copy-by-copy}), we record, for each copy and within a single time bin $\Delta\tau$, both the coincidence counts and the total detected photon number.
This method has two main advantages. First, the pulse-to-pulse photon-number fluctuations are already suppressed at this photon-number level. Second, because each copy is analyzed individually, the influence of any remaining number fluctuations can be accurately controlled.

{\it Method 2: Post-selection}.--- In the second method, the total photon number and the coincidences for each copy are stored electronically. Since all copies satisfy $g^{(2)}(0)<1$, we can further reduce the pulse-to-pulse number fluctuations by post-selection. Specifically, in evaluating Eq.~(\ref{copy-by-copy}), we include only those copies whose $\langle \hat{n}_i( \hat{\rho}_{b'} )\rangle$ lies within a very narrow range.

{\it Method 3: Attenuation and single-photon detectors}.--- The third method is \textit{indeed}  \textit{original} and  uses standard single-photon detectors. The number fluctuations are reduced by several orders of magnitude. The counter-displaced states $\hat{\rho}_{b'}$ are sent through an absorptive medium. This process is modeled as a beam-splitter transformation,  $\hat{b}''_{\rm \scriptscriptstyle }=\sqrt{\eta}\, \hat{b}'+ \sqrt{1-\eta} \, \hat{b}_{\rm vac}$~\cite{tan2008quantum,fesquet2023perspectives,huck2009demonstration,di2012quantum}, where $\eta$ is the transmission coefficient.

As shown experimentally and analytically to first order in Ref.~\cite{di2012quantum}, the value of $g^{(2)}(0)$ remains unchanged when a state passes through an amplitude-decaying medium, even though the mean photon number is strongly reduced. An exact analytical proof of the invariance of $g^{(2)}(0)$ is provided in Appendix~\ref{sec:appendix_attenuation}. Such an attenuating device is placed between the  $\hat{\rho}_{b'}$ mode and the $g^{(2)}(\tau)$ histogram in Fig.~\ref{fig1}.
Although the transformation $\hat{b}''_{\rm \scriptscriptstyle }=\sqrt{\eta}\, \hat{b}'+ \sqrt{1-\eta} \, \hat{b}_{\rm vac}$ rapidly degrades quadrature squeezing, it preserves the normalized nonclassicality measure $g^{(2)}(0)$. This occurs because, at room temperature, the occupation of optical vacuum modes ($\hat{b}_{\rm vac}$) is negligible. At lower frequencies, a cooled medium is required.



When the intensity becomes very small, the total number of detected events may be insufficient to measure $g^{(2)}(0)$. For this reason, experiments such as Ref.~\cite{di2012quantum,tame2013quantum} use a heralded single-photon source based on parametric down-conversion. One photon is sent through the absorptive medium, while the other is used to herald the photon generation time and suppress dark counts.

Using this approach, a state $\hat{\rho}_{b'}$ containing about 20 photons per pulse can be attenuated to an average of 0.1 photon per pulse (denoted $\hat{\rho}_{b''}$) without changing the value of $g^{(2)}(0)$. This photon number is typical in single-photon $g^{(2)}(0)$ measurements, and the relative pulse-to-pulse number fluctuations are reduced to about 1/200 for the parameters assumed here. Such pulse-to-pulse photon-number distributions are common in experiments with pulsed sources~\cite{loudon2000quantum,fox2006quantum,migdall2013single,lounis2005single}.

\section{Summary and Conclusions} \label{sec:conclusions}

We raise a fundamental question whose answer is both scientifically intriguing and practically important for performance benchmarking in some continuous-variable QKD protocols, long-distance quantum information distribution, and quantum-enhanced imaging: Can one determine whether a state is quadrature squeezed when the amplitudes (displacements) of the collected copies are random?

Such a situation naturally occurs when initially identical squeezed copies propagate through a dynamical channel. Examples include microwave- or optical-frequency quantum communication between mobile platforms, where amplitude loss or squeezing degradation fluctuates in time and therefore varies from copy to copy.

We approach this problem from a different perspective. First, we engineer an interaction Hamiltonian that converts quadrature squeezing into number squeezing. We show that the interaction $\hat{B}_n(\theta)$ does not generate nonclassicality unless the incident copies already contain it. When the copies are randomly displaced, neither quadrature squeezing nor number squeezing can be directly witnessed, since there is no fixed mean value around which noise statistics can be evaluated. Nevertheless, the presence of nonclassicality can still be inferred from the statistics of the second-order correlation function $g^{(2)}(\tau)$ using a histogram-based measurement.

A sub-Poissonian value $g^{(2)}(0)<1$ can be observed as long as each copy individually satisfies $g^{(2)}(0)<1$. To improve the visibility of this signature, we propose several techniques that reduce the value of $g^{(2)}(0)$ and narrow the pulse-to-pulse photon-number distribution before constructing the $g^{(2)}(\tau)$ histogram. These methods enhance the robustness and reliability of detecting sub-Poissonian statistics. Moreover, the detection of $g^{(2)}(0)$ can be implemented using conventional single-photon detection procedures commonly employed in experiments~\cite{loudon2000quantum,fox2006quantum,migdall2013single,lounis2005single}.


Specifically, we propose the following approaches. First, a counter-displacement operation $\hat{D}(\alpha')$ can be applied to reduce the value of $g^{(2)}(0)$ further below unity (Fig.~\ref{fig3}b). Second, a post-selection strategy can be employed to suppress pulse-to-pulse photon-number fluctuations among different copies, even though fluctuations within individual displacements are already reduced. Third, we introduce an \textit{elegant} method that is compatible with standard single-photon detection schemes, where the mean photon number per pulse is on the order of $0.1$. This method exploits the fact that attenuation of a signal $\hat{\rho}_{b'}$ does not change its $g^{(2)}(0)$ value. This invariance has been demonstrated experimentally (see Refs.~\cite{di2012quantum,tame2013quantum}) and is derived analytically in Appendix~\ref{sec:appendix_attenuation}. Importantly, attenuation significantly narrows the pulse-to-pulse photon-number fluctuations and makes standard methods for single-photon detection available.

The operation $\hat{B}_n(\theta)$ is formally analogous to a standard beam splitter, which redistributes quadrature squeezing between two modes. The essential difference is that, while an ordinary beam splitter can only transfer quadrature squeezing from one mode to another, the operator $\hat{B}_n(\theta)$ converts quadrature squeezing in the first mode into number squeezing in the second mode. This behavior arises because the usual annihilation operator $\hat{b}=(\hat{x}_b+i\hat{p}_b)/\sqrt{2}$ in the standard beam-splitter Hamiltonian $\hat{a}^\dagger\hat{b}+\mathrm{H.c.}$ is replaced by the operator $\hat{b}_n=(\hat{n}_b+i\gamma_0\hat{\Phi}_b)/\sqrt{2\gamma_0}$, whose conjugate variables are the photon-number operator $\hat{n}_b$ and the phase operator $\hat{\Phi}_b$~\cite{vaccaro1990physical,tasgin2019anatomy,pegg1989phase,barnett1989hermitian,barnett2007quantum}.

The main goal of this work is to determine whether squeezing in randomly displaced copies of a squeezed state can be identified, at least in principle. This question is particularly relevant for quantum-secure and insecure~\cite{tasgin2025encoding} mobile communication. In stable channels, such as fiber-optic links, amplitude loss can be accurately estimated. In contrast, when communication takes place through the atmosphere—for example, between moving vehicles—the losses fluctuate due to atmospheric dynamics, leading to random variations in the amplitudes of the received copies. In such scenarios, the method introduced here becomes essential.

Finally, in Appendix~\ref{sec:appendix_Hamiltonian}, we discuss several physical platforms in which the interaction Hamiltonian implementing the $\hat{B}_n(\theta)$ transformation can be realized. In Appendix~\ref{sec:appendix_attenuation}, we also provide an exact analytical proof that the value of $g^{(2)}(0)$ remains invariant for optical signals propagating through an attenuating medium.

\vspace{3 cm}

\appendix

\section{$\hat{B}_n(\theta)$ Hamiltonian} \label{sec:appendix_Hamiltonian}

\subsection{Circuit QEC and Microwave}

The operator $\hat{b}_n$ can be naturally realized in circuit quantum electrodynamics (cQED)~\cite{barzanjeh2011entangling,vion2002manipulating,blais2021circuit}. In particular, the Hamiltonian $\hat{B}_n(\theta)$ can be implemented through the interaction of superconducting circuits with microwave drives~\cite{blais2021circuit}.

In the harmonic (small-fluctuation) regime, i.e., when $E_{\rm \scriptscriptstyle J} \gg E_{\rm \scriptscriptstyle C}$, the Josephson potential of a Cooper-pair box (CPB) can be linearized. This yields the Hamiltonian
\begin{equation}
	{\cal \hat{H}}_{\rm \scriptscriptstyle J}
	= E_{\rm \scriptscriptstyle C}\hat{n}^{2}
	+ \frac{E_{\rm \scriptscriptstyle J}}{2}\hat{\Phi}^{2},
	\label{HJ}
\end{equation}
where $\hat{n}$ is the Cooper-pair number operator and $\hat{\Phi}$ is the superconducting phase difference across the junction, satisfying $[\hat{n},\hat{\Phi}]=i$. Here, $E_{\rm \scriptscriptstyle C}$ denotes the charging energy and $E_{\rm \scriptscriptstyle J}$ the Josephson energy.

The plasma-mode operator
\begin{equation}
	\hat{b} = \frac{c_1 \hat{n} + i\, c_2 \hat{\Phi}}{\sqrt{2}}
	\label{plasma_operator}
\end{equation}
diagonalizes the linearized Hamiltonian~(\ref{HJ}). The coefficients are given by
\begin{equation}
	c_1=2\left( \frac{2E_{\rm \scriptscriptstyle C}}{E_{\rm \scriptscriptstyle J}} \right)^{1/4},
	\qquad
	c_2=2\left( \frac{E_{\rm \scriptscriptstyle J}}{2E_{\rm \scriptscriptstyle C}} \right)^{1/4}.
\end{equation}
Thus, the operator $\hat{b}_n$ used throughout this paper appears naturally in cQED as the standard plasma operator of a Josephson junction~\cite{blais2021circuit}.

Using this operator, the capacitive coupling term proportional to $\hat{n}(\hat{a}^\dagger+\hat{a})$ takes the form
\begin{equation}
	\hat{H}_{\rm cav\text{--}{\rm \scriptscriptstyle J}}
	= \hbar \xi (\hat{b}^\dagger \hat{a} + \hat{a}^\dagger \hat{b}),
	\label{HcavJ}
\end{equation}
after substituting Eq.~(\ref{plasma_operator}) and applying the rotating-wave approximation~\cite{blais2021circuit}. This Hamiltonian is exactly the $\hat{B}_n(\theta)$ interaction introduced in Eq.~(\ref{Bn_operator}).

Consequently, if the incident cavity mode $\hat{a}$ is quadrature-squeezed, this interaction transfers the squeezing into reduced uncertainty of the Cooper-pair number. In other words, nonclassicality in the form of quadrature squeezing is converted into number squeezing of the CPB.

We present this example to illustrate that the operators $\hat{b}_n$ and $\hat{B}_n(\theta)$ arise naturally in cQED systems, which have recently attracted significant interest. To use such a system for detecting nonclassicality in randomly displaced copies of squeezed light, one must either (i) directly measure the fluctuations of the Cooper-pair number, or (ii) transfer the excitation to another microwave cavity where $g^{(2)}(\tau)$ can be measured.

The latter option requires a beam-splitter–like interaction,
\begin{equation}
	{\cal \hat{H}}
	= \hbar \chi (\hat{c}^\dagger \hat{J} + \text{H.c.}),
\end{equation}
where $\hat{J}$ lowers the number of Cooper pairs by one. Such interactions arise from inelastic Cooper-pair tunneling under a DC bias~\cite{wood2021josephson,leppakangas2006tunneling}. However, multi-photon emission into a single cavity via this mechanism has not yet been experimentally demonstrated. Existing studies focus on single-photon emission in the large-$E_{\rm C}$ regime~\cite{leppakangas2015antibunched} or on generating entanglement between photons emitted into separate cavities~\cite{wood2021josephson}. We also note that the operator $\hat{J}$ does not behave as a standard bosonic annihilation operator $\hat{f}$ satisfying $\hat{n}=\hat{f}^\dagger\hat{f}$.

Finally, we emphasize that the commutation relation $[\hat{n},\hat{\Phi}]=i$ holds generally for a Cooper-pair box.

\subsection{Optical regime using atomic ensembles}

The Hamiltonian $\hat{B}_n(\theta)$ can also be engineered through the interaction of a single-mode optical field $\hat{a}$ with an ensemble of bosonic atoms. To obtain full control over the effective interaction, we consider a four-level atomic system with two ground states $|g_1\rangle$, $|g_2\rangle$ and two excited states $|e_1\rangle$, $|e_2\rangle$.
The optical mode $\hat{a}$ couples to the transitions $|g_1\rangle \leftrightarrow |e_1\rangle$ and $|g_2\rangle \leftrightarrow |e_2\rangle$⟩, with detunings $\Delta_1$ and $\Delta_2$,  respectively. These detunings are assumed to be large so that the excited states can be adiabatically eliminated.

To gain additional tunability, we apply two classical Raman control beams with complex Rabi frequencies $\Omega_1 e^{i\phi_1}$ and $\Omega_2 e^{i\phi_2}$, detuned from the corresponding optical transitions by  $\Delta_{\rm ctrl}^{(1)}$ and $\Delta_{\rm ctrl}^{(2)}$. 
In addition, a microwave field
\begin{equation}
	\frac{\hbar\Omega_{\rm mw}}{2} ( e^{i\theta_{\rm mw}} |g_1\rangle\langle g_2| + H.c. ),
\end{equation}
coherently couples the two ground states and allows control of the collective spin orientation~\cite{MandelWolf,saffman2009spin,GaoAVS2023,schleier2011cavity,hammerer2010quantum,Xue2022Controlling}.

The atomic ensemble is described using bosonic annihilation and creation operators $\hat{b}_1$, $\hat{b}_1^\dagger$ and $\hat{b}_2$, $\hat{b}_2^\dagger$, corresponding to populations in states $|g_1\rangle$ and $|g_2\rangle$. 
The collective spin operators are
\begin{equation}
	\hat{S}_+=\hat{b}_1^\dagger \hat{b}_2, \quad \hat{S}_z=\frac{1}{2}\left(\hat{b}_1^\dagger \hat{b}_1-\hat{b}_2^\dagger \hat{b}_2 \right)
\end{equation}
and the population-difference (number) operator is
\begin{equation}
	\hat{n}_b=\hat{n}_1-\hat{n}_2, \quad \hat{n}_j=\hat{b}_j^\dagger \hat{b}_j .
\end{equation}
Because all optical couplings are far detuned, the populations of the excited states remain negligible, and the total number of atoms $N$ is conserved within the two ground states.

After adiabatic elimination of the excited states, two effective interactions emerge between the optical mode $\hat{a}$ and the ground-state manifold. The first is a dispersive (AC Stark shift) interaction,
\begin{equation}
	\hat{\mathcal H}_{\rm disp}
	= -\hbar\, \hat{a}^\dagger \hat{a}
	\left(
	\frac{|g_1|^2}{\Delta_1}\,\hat{n}_1
	+
	\frac{|g_2|^2}{\Delta_2}\,\hat{n}_2
	\right),
	\label{Hdisp}
\end{equation}
where $g_1$ and $g_2$ are the single-photon coupling strengths. The second is a Raman-induced spin-flip interaction,
\begin{equation}
	\hat{\mathcal H}_{\rm Ram}
	= -\hbar
	\left(
	\frac{g_1 \Omega_2^{*}}{\Delta_{\rm eff}} \,
	\hat{a}^\dagger \hat{S}_{-}
	+
	\frac{g_2 \Omega_1^{*}}{\Delta_{\rm eff}} \,
	\hat{a}^\dagger \hat{S}_{+}
	+ \text{H.c.}
	\right),
	\label{HRam}
\end{equation}
where $\Delta_{\rm eff}$ is the effective two-photon detuning and $\hat{S}_-=\hat{S}_+^\dagger$.

To obtain the desired form of the interaction, we operate in the large classical-amplitude regime, writing the optical field as $\hat{a}=\alpha+ \delta \hat{a}$, with 
$|\alpha|\gg (\delta \hat{a})_{\rm fluc}$, as commonly used in quantum optics~\cite{vitali2007optomechanical,CGenes08,navarrete2017general,aspelmeyer2014cavity}. The photon-number operator then becomes
\begin{equation}
	\hat{a}^\dagger \hat{a} = |\alpha|^2 + \alpha^*\, \delta\hat{a} + \alpha \, \delta\hat{a}^\dagger + \delta\hat{a}^\dagger \, \delta \hat{a},
\end{equation}
where the last term is second order in fluctuations and can be neglected away from criticality~\cite{aspelmeyer2014cavity,lambert2004entanglement,navarrete2017general}.

Combining the dispersive and Raman contributions, the effective interaction Hamiltonian takes the form
\begin{equation}
		\mathcal H_{\rm int}
		= -\hbar\, \hat{a}^\dagger \left[
		a_d\, \hat{n}_b
		+ \beta\, \hat{S}_-
		\right] + \text{H.c.},
		\label{H_final}
\end{equation}
with
\begin{equation}
a_d \: \propto \alpha \left(\frac{|g_1|^2}{\Delta_1} - \frac{|g_2|^2}{\Delta_2} \right) \quad  {\rm and}  \quad 
\beta \propto \left( \frac{g_1 \Omega_2^*}{\Delta_1}  - \frac{g_2 \Omega_1^*}{\Delta_2} \right).
\end{equation}
In the regime of small collective excitations, corresponding to small angular deviations of the collective spin, the spin-lowering operator may be approximated as~\cite{saffman2009spin,GaoAVS2023,schleier2011cavity,hammerer2010quantum,Xue2022Controlling}
\begin{equation}
	\hat{S}_- \simeq \sqrt{N}\left( 1 - i\hat{\Phi}_b + \ldots \right),
\end{equation}
where $\hat{\Phi}_b$ denotes the small phase fluctuation. In this limit, the approximate canonical commutation relation $[\hat{n}_b,\Phi_b]\simeq i$ holds. Substituting this expansion into the interaction Hamiltonian yields
\begin{equation}
	{\mathcal H}_{\rm int}
	= -\hbar \, \hat{a}^\dagger \left[
	a_d \,\hat{n}_b
	+ i \beta'\, \hat{\Phi}_b
	\right] + \text{H.c.},
\end{equation}
with $\beta'=\sqrt{N} \, \beta$. This Hamiltonian has exactly the structure required for the $\hat{B}_n(\theta)$ interaction, since $\hat{n}_b$ and $\hat{\Phi}_b$ form an approximately canonical pair.

\emph{Transfer of squeezing.—}
With this interaction, quadrature squeezing in the optical field $\hat{a}$ is converted into number squeezing in the ground-state doublet of the atomic ensemble.

Using the Holstein–Primakoff mapping~\cite{lambert2005entanglement},
\begin{equation}
\hat{S}_+= \sqrt{N-\hat{c}^\dagger \hat{c}\,} \, \hat{c} \quad {\rm and} \quad \hat{S}_z = N - \hat{c}^\dagger \hat{c},
\end{equation}
the resulting quasiparticle mode $\hat{c}$ (or an equivalent bosonic mode) inherits this number squeezing.

Finally, the number squeezing stored in the ensemble can be transferred to another optical mode $\hat{d}$ by coupling the two ground states to a single excited state using $\hat{d}$ and a classical control field, as demonstrated in Refs.~\cite{hammerer2010quantum,tacsgin2011spin}. The resulting beam-splitter–type interaction 
($\hat{S}_+ \hat{d}+H.c.$ ) enables faithful transfer of spin squeezing to the optical field~\cite{kuzmich1997spin,drummond2004quantum,hammerer2010quantum}.

\subsection{Nonlinear optics}

One can also obtain a similar Hamiltonian using nonlinear optical components, such as optical parametric amplifiers that rely on Kerr or other nonlinearities. When these devices operate above threshold—i.e., in the limit-cycle regime~\cite{navarrete2017general}—the optical annihilation operators can be linearized around stable bright states. By tuning the relative strengths of the nonlinear interactions, one can engineer an effective Hamiltonian that corresponds to the operator $\hat{B}_n(\theta)$.

\section{Exact derivation of the invariance of $\mathbf{g^{(2)}(0)}$ under attenuation} \label{sec:appendix_attenuation}

When a signal passes through an absorbing medium, its amplitude is attenuated. Nonclassical features are also known to deteriorate according to the transformation~\cite{tan2008quantum,fesquet2023perspectives,huck2009demonstration,di2012quantum}
\begin{equation}
	\hat{a}{\rm \scriptscriptstyle T}= \sqrt{\eta} \, \hat{a} + \sqrt{1-\eta} \, \hat{b}_{\rm vac},
	\label{absorption_transformation}
\end{equation}
where $\eta$ is the transmission coefficient, $\hat{b}_{\rm vac}$ denotes a vacuum mode, and $\hat{a}{\rm \scriptscriptstyle T}$ is the transmitted mode.

At optical frequencies and room temperature, the occupation of the environmental modes can be taken to be zero. This assumption applies to optical quantum states propagating through the atmosphere~\cite{tan2008quantum,fesquet2023perspectives}, to surface plasmon polaritons (SPPs) propagating along absorptive metallic nanowires~\cite{huck2009demonstration,di2012quantum}, and to the controlled attenuating device placed between $\hat{\rho_{b'}}$ state and the $g^{(2)}(\tau)$ histogram.


Quadrature squeezing is highly sensitive to attenuation, and the degree of squeezing deteriorates proportionally to the signal loss. Indeed, Eq.~(\ref{absorption_transformation}) shows that the noise properties of an initially squeezed mode $\hat{a}$ rapidly approach those of the vacuum (i.e., no squeezing) in the transmitted mode $\hat{a}_{\rm \scriptscriptstyle T}$ as $\eta$ decreases.

The second-order correlation function,
\begin{equation}
	g^{(2)}(0) =\frac{ \langle \hat{a}^\dagger \hat{a}^\dagger \hat{a \hat{a}} \rangle }{(\langle \hat{a}^\dagger \hat{a} \rangle)^2},
\end{equation}
however, is a normalized quantity. As a result, when the field operator $\hat{a}$ is attenuated by a factor $\eta$, the value of $g^{(2)}(0)$ remains unchanged. Reference~\cite{di2012quantum,tame2013quantum} has already demonstrated this invariance experimentally and within a first-order analytical approximation, showing that $g^{(2)}(0)$ remains constant even after transmission through long metallic stripes with transmission coefficients as small as $\eta \sim 10^{-4}$~\cite{di2012quantum}.

Here, we provide an exact analytical proof, without relying on a first-order approximation, using the transformation in Eq.~(\ref{absorption_transformation}). This result~\cite{tame2013quantum,lopaeva2013experimental,sekatski2012detector} is directly relevant to our $g^{(2)}(0)$ measurement protocol (method~3 in Sec.~\ref{sec:histogram}). The expectation values of the transmitted photon number and its square are
\begin{equation}
	\langle \hat{n}_{\rm \scriptscriptstyle T} \rangle = \eta \, \langle \hat{n}_a\rangle + (1-\eta) \, \langle \hat{n}_b \rangle,
\end{equation}
and
\begin{eqnarray}
	\langle \hat{n}{\rm \scriptscriptstyle T}^2 \rangle &=&
	\eta^2 \, \langle \hat{n}_a^2\rangle + (1-\eta)^2 \,\langle\hat{n}_b^2\rangle + 2\eta(1-\eta) \, \langle\hat{n}_a\rangle\langle\hat{n}_b\rangle \qquad
	\label{nT}
	\\
	&+& \eta(1-\eta) \left[ \langle \hat{n}_a+1\rangle \langle \hat{n}_b\rangle + \langle \hat{n}_b+1\rangle \langle \hat{n}_a\rangle \right].
	\label{nt2}
\end{eqnarray}
Here we have used the fact that expectation values containing operators such as $\hat{b}$ or $\hat{b}^2$ (i.e., without corresponding creation operators) vanish for the vacuum state $|0\rangle_b$. Consequently, both $\langle \hat{n}_b\rangle$ and $\langle \hat{n}_b^2\rangle$ vanish in Eqs.~(\ref{nT}) and~(\ref{nt2}).

The second-order correlation function of the transmitted mode then reads
\begin{eqnarray}
	g_{\rm \scriptscriptstyle T}^{(2)}(0)
	&=& \frac{\langle \hat{n}{\rm \scriptscriptstyle T}^2\rangle - \langle \hat{n}{\rm \scriptscriptstyle T}\rangle}{\langle \hat{n}_{\rm \scriptscriptstyle T}\rangle^2}
	\\
	&=& \frac{\eta^2 \langle \hat{n}_a^2 \rangle - \eta^2 \langle \hat{n}_a \rangle}{\eta^2 \langle \hat{n}_a \rangle^2}
	= g_a^{(2)}(0).
\end{eqnarray}
Thus, $g^{(2)}(0)$ remains invariant under attenuation. This property also holds for other normally ordered operators when the occupation of the vacuum modes is negligible~\cite{lopaeva2013experimental,sekatski2012detector}.

Consequently, in method~3 of Sec.~\ref{sec:histogram}, when an initially $\sim 20$–photon-occupied state $\hat{\rho}_{b'}$ is attenuated in an absorbing medium, the resulting state $\hat{\rho}_{b''}$—with a mean photon number as low as $0.1$ per pulse~\cite{loudon2000quantum,fox2006quantum,migdall2013single,lounis2005single}—exhibits the same value of $g^{(2)}(0)$. At the same time, the pulse-to-pulse photon-number fluctuations among the received copies originating from the initial state $\hat{\rho}_t$ are suppressed by a factor of approximately $200$.

\bibliography{bibliography}	

\end{document}